\documentclass{PoS}
\usepackage{graphicx}
\usepackage{subfigure}
\usepackage{float}

\title{Studying photon structure at an EIC}

\ShortTitle{Photon structure at an EIC}

\author{\speaker{X. Chu}\\
        Key Laboratory of Quark and Lepton Physics (MOE)  and \\Insititute of Particle Physics, Central China Normal University, \\Wuhan 430079, China\\
        Physics Derpartment, Brookhaven National Laboratory, \\Upton, New York 11973, USA \\
        E-mail: \email{xchu@mails.ccnu.edu.cn}}
\author{E.C. Aschenauer\\
        Physics Derpartment, Brookhaven National Laboratory, \\Upton, New York 11973, USA \\
        E-mail: \email{elke@bnl.gov}}
\author{J.H. Lee\\
        Physics Derpartment, Brookhaven National Laboratory, \\Upton, New York 11973, USA \\
        E-mail: \email{jhlee@bnl.gov}}

\abstract{A future Electron-Ion Collider (EIC) will deliver luminosities of $10^{33} - 10^{34}$ cm$^{-2}$s$^{-1}$ for collisions of polarized electrons and protons and heavy ions over a wide range of center-of-mass energies (40 $\mathrm{GeV}$ to 145 $\mathrm{GeV}$).
One of its promising physics programs is to study the partonic structure of quasi-real photons. 
Measuring di-jet in photoproduction events, one can effectively access the underlying parton dynamics of the photons through the selection of the resolved photon processes.
In this paper, we discuss the feasibility of tagging resolved photon processes and measuring the di-jet cross section as a function of jet transverse momentum in ranges of $x_{\gamma}^{rec}$ at an EIC.
These studies show that parton distributions in the photon can be extracted at an EIC.}

\FullConference{XXIV International Workshop on Deep-Inelastic Scattering and Related Subjects\\
		11-15 April, 2016\\
		DESY Hamburg, Germany}

\begin{document}

\section{Introduction}
The studies of hadronic photon structure have a long history since the theoretical work in 1971~\cite{modelcal1}. On the experimental side, in the past decades we have seen large progress due to the data from HERA, where high energy electrons were collided with protons. The Parton Distribution Functions (PDFs) of the unpolarized photon can be extracted by tagging high transverse energy ($E_{t}$) jets ~\cite{modelcal2,modelcal3}, high-$p_{T}$ charged particles~\cite{modelcal4} or heavy quarks in deep-inelastic scattering (DIS) at HERA, although these measurements have relatively large uncertainties. The future EIC, which will deliver luminosities of $10^{33} - 10^{34}$ cm$^{-2}$s$^{-1}$ for polarized electron and proton collisions~\cite{modelcal5,modelcal6}, is an excellent tool to study the partonic photon structure with high precision. More importantly, one can access the unknown polarized photon PDFs. 

\par In $ep$ collision, the interaction of electrons and protons at low virtuality is dominated by photoproduction processes with electrons scattering at small angles. Such reactions proceed via two classes of processes. The first one is a direct process (for example the photon-gluon fusion process in Fig. 1a), in which the photon couples directly to a parton from the proton with the entire energy of the photon going into the hard scattering. The second one is a resolved process (Fig. 1b), the photon manifests a hadronic structure because of quantum fluctuations, then partons from the photon and proton interact via the strong interaction. These resolved events allow for the study of the partonic structure of the photon.

\begin{figure}[H]
\centering
\subfigure[] {\includegraphics[height=0.33\textwidth]{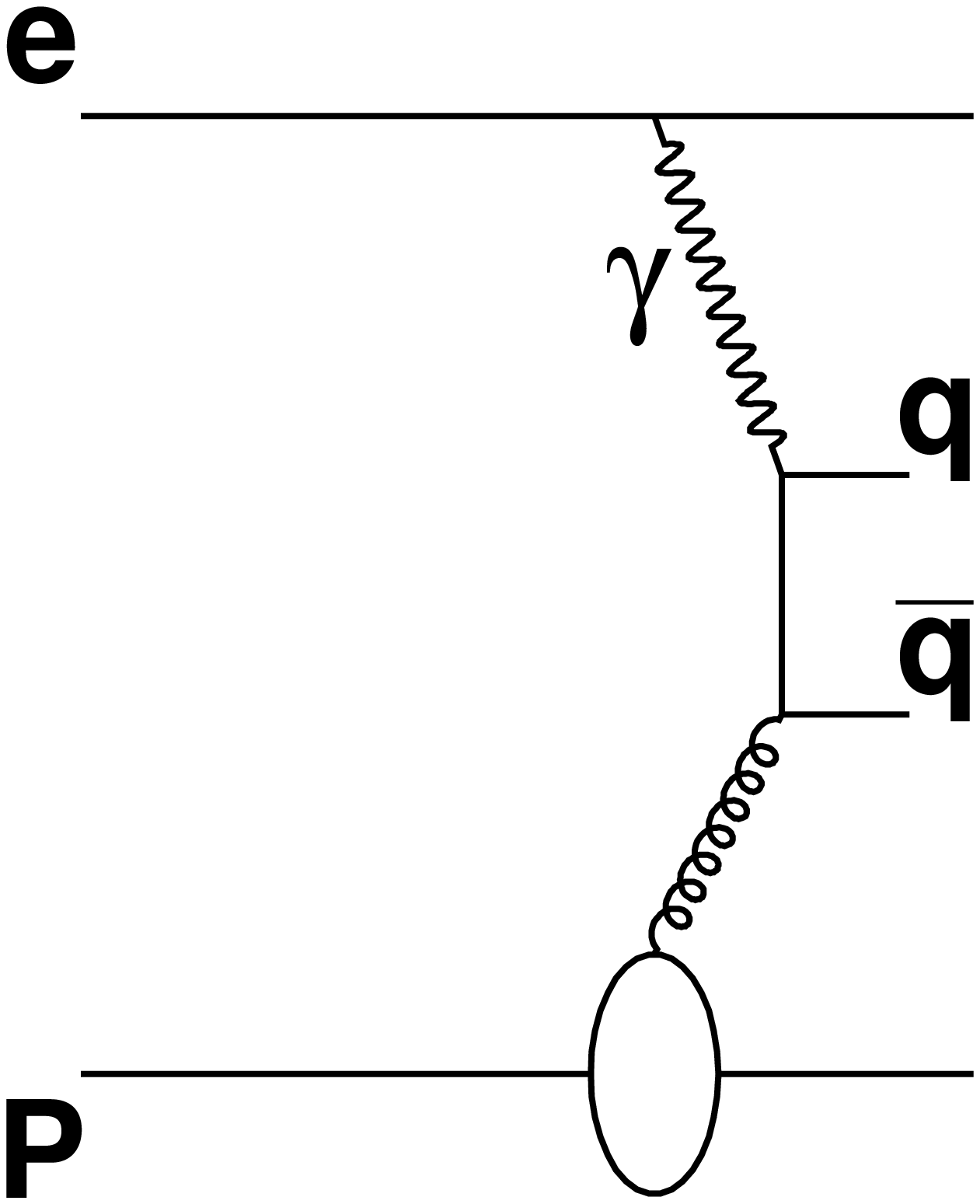}}
\subfigure[] {\includegraphics[height=0.33\textwidth]{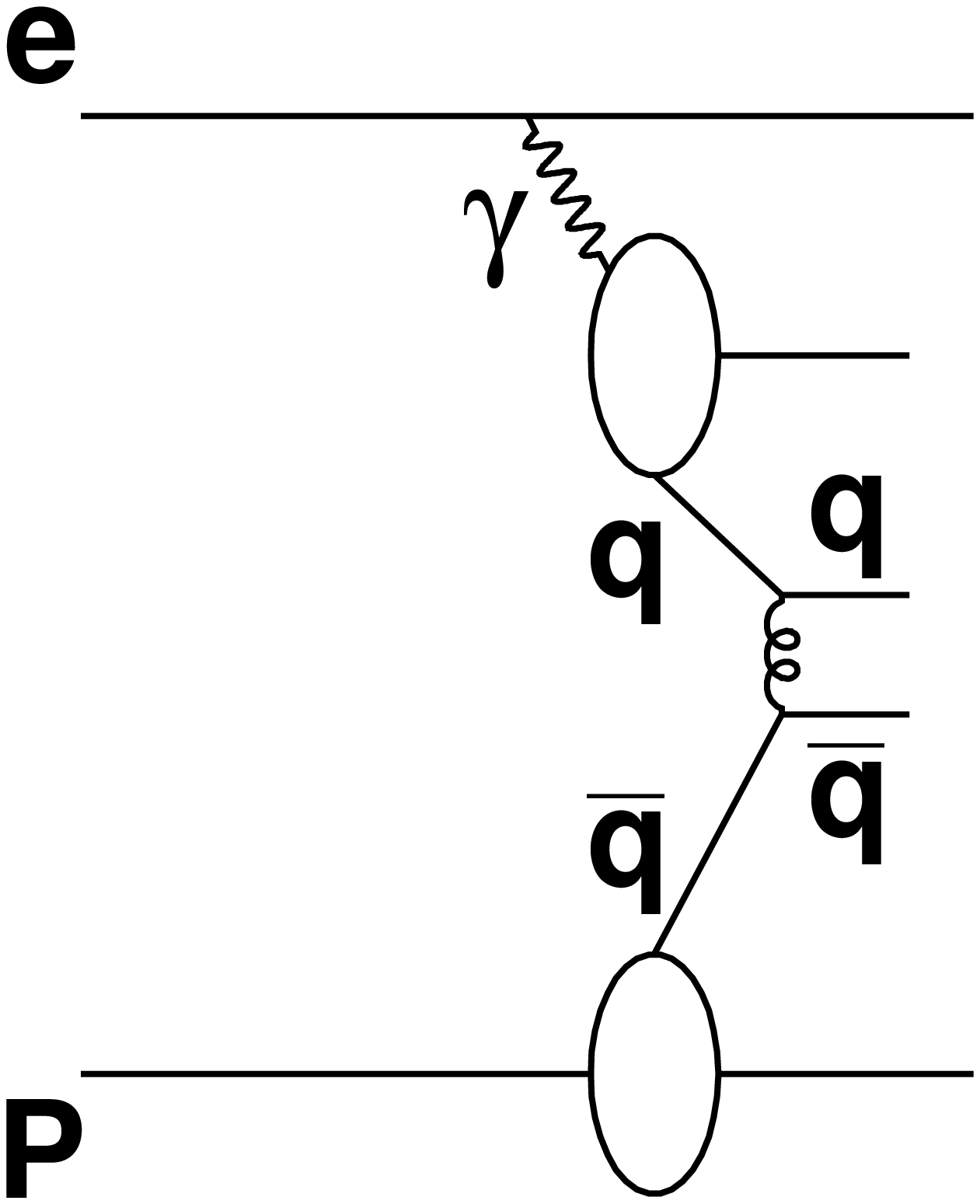}}
\caption{Examples of diagrams for direct (a) and resolved (b) processes in electron-proton scattering.}
\label{}
\end{figure}

\par These proceedings are organized as follows: In Sec. 2, we present a method of reconstructing $x_{\gamma}$. The Monte Carol simulations at a future EIC of the di-jet cross section is validated by the data collected with H1 detector at HERA in Sec. 3. Then we will briefly discuss the method of distinguishing resolved and direct processes, and the extraction of photon PDFs at EIC. Finally we close with a summary.

\section{Method of reconstructing $x_{\gamma}$}
Predictions for the di-jet cross section are usually obtained in leading order quantum chromodynamics (LO QCD) by convoluting the parton densities in the photon and those in proton. As the parton densities in the proton are well known, a measurement of the di-jet cross section can be used to extract information on the parton densities of the photon~\cite{modelcal7,modelcal8}. In order to extract the photon structure information through di-jet production, first we need to select resolved processes. The best variable to separate the two types of processes is $x_{\gamma}$, which represents the momentum fraction of the parton in the photon. Since in the direct processes the photon interacts with a parton from the proton as a structureless particle, $x_{\gamma}$ of direct processes is equal to 1 theoretically. Whereas in the resolved processes, the photon behaves like a source of partons, with only a fraction of its momentum participating in the hard scattering, therefore, the corresponding $x_{\gamma}$ should be smaller than 1. 
The variable $x_{\gamma}$ can be reconstructed knowing the momentum and angles of di-jet as,\\
\begin{equation}
x_{\gamma}^{rec} =  \frac{1}{2E_{e}y} (p_{T,1}e^{-\eta_{1}}+p_{T,2}e^{-\eta_{2}}),
\end{equation}
where $E_{e}$ is the electron energy and $y$ is the energy fraction taken by the photon from the electron ($y=\frac{E_{\gamma}}{E_{e}}$). Eq. (2.1)  is valid in the lab frame.

\section{Verification of Simulation with HERA data}
\begin{figure}[H]
  \centering
  \includegraphics[width=0.4\textwidth]{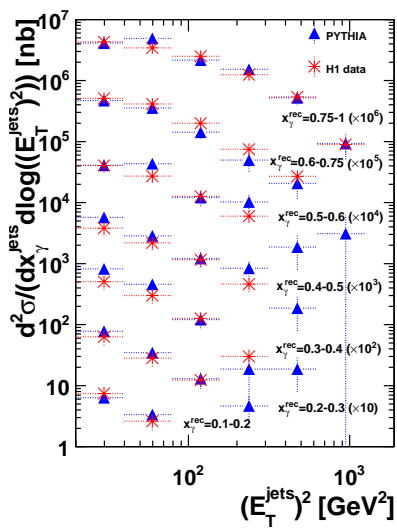}
  \caption{Comparison of the di-jet cross section from PYTHIA simulation with the selected HERA data. We used the same parameters as the data in the simulation.}
\end{figure}
The di-jet cross section measured by H1 at HERA~\cite{modelcal7} is shown in Fig. 2 as a function of the squared jet transverse energy $E_{T}^{jets}$ in ranges of reconstructed $x_{\gamma}$. Here we use $E_{T}^{jets}$ to represent the average transverse energy of trigger jet and associated jet, $E_{T}^{jets}$ is required to be above $E_{T}^{jets} = 10$ $ \mathrm{GeV}$. The ratio of the difference and the sum of the transfers energies of the jets is required to be in the range of  $\frac{|E_{T}^{jet1}-E_{T}^{jet2}|}{(E_{T}^{jet1}+E_{T}^{jet2})}<0.25$. The fractional photon energy is between $0.2<y<0.83$. The average of pseudo-rapidity of the two jets is restricted between $0 < \frac{\eta^{jet1}+\eta^{jet2}}{2} <2$, and the difference of the jet pseudo-rapidities is required to be within $|\Delta\eta^{jets}| < 1$. The simulation results are obtained with 27 $\mathrm{GeV}$ electrons colliding on protons of 820 $\mathrm{GeV}$, and the comparison shows the simulation reproduces the measured data reasonably well.

\section{Photon structure at EIC}
In this study, jets are reconstructed with the anti-$k_{T}$ algorithm~\cite{modelcal9}, which is parametrised by the power of the evergy scale in the distance measure. All the stable and visible particles produced in the collision with $p_{T}>250$ MeV$/c$ and $-4.5<\eta<4.5$ in the laboratory system are taken as input. 

The resolution parameter, $R$, is set to 1.
This simulation for EIC is obtained colliding 20 $\mathrm{GeV}$ electrons with 250 $\mathrm{GeV}$ protons. Jets with the highest $p_{T}$ are called the trigger jet, then we selected the jet with second highest $p_{T}$ to be the associate jet. Events are selected with the requirement that the trigger jet has $p_{T}^{jet}> 5$ $\mathrm{GeV}$ and the associated jet has $p_{T}^{jet}> 4.5$ $\mathrm{GeV}$. The PYTHIA 6.4 event generator for electron-proton interactions is used. The photoproduction events are selected by requiring the four-momentum of the photon of $Q^{2}< 1$ $\mathrm{GeV^{2}}$. The inelasticity cut is $0.01<y<0.95$. The event kinematic variables $Q^{2}$ and $y$ are obtained directly from PYTHIA simulations without applying the event kinematics reconstruction method.
\par We reconstruct $x_{\gamma}$ with di-jet events. The strong correlation between the reconstructed $x_{\gamma}$ and the input $x_{\gamma}$ from PYTHIA is shown in Fig. 3(a), and it indicates the di-jet observable is ideal for this measuremet. The $x_{\gamma}^{rec}$ distribution is shown in Fig. 3(b). The resolved (direct) process dominates at low (high) $x_{\gamma}^{rec}$ regime, which provides a good way of separating the two types of processes. For example, by selecting events with $x_{\gamma}^{rec} < 0.6$ or $> 0.6$, one can divide the di-jet events into subsamples in which the resolved and direct processes dominate.

\begin{figure}[H]
\centering
\subfigure[] {\includegraphics[height=0.33\textwidth]{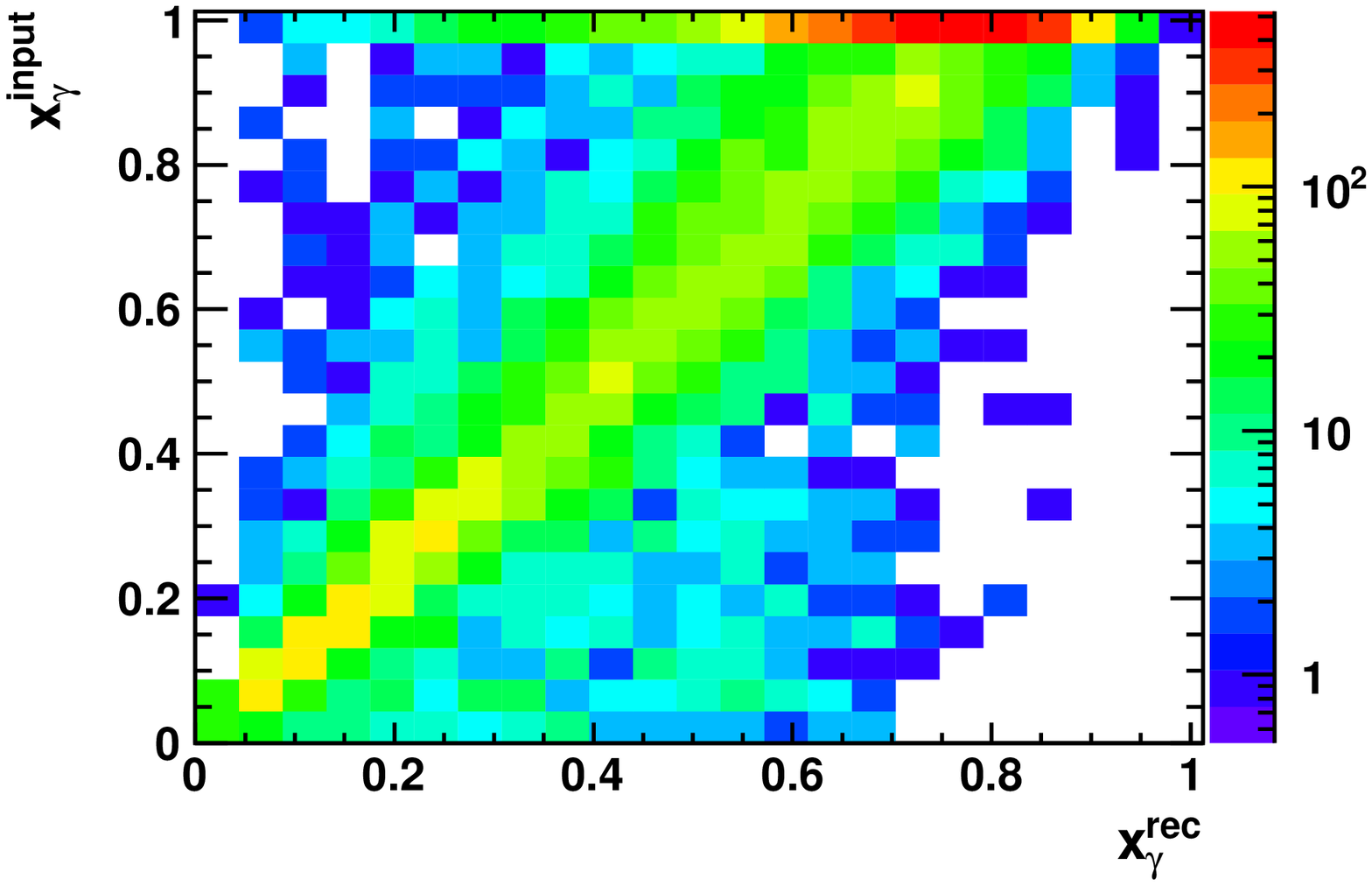}}
\subfigure[] {\includegraphics[height=0.33\textwidth]{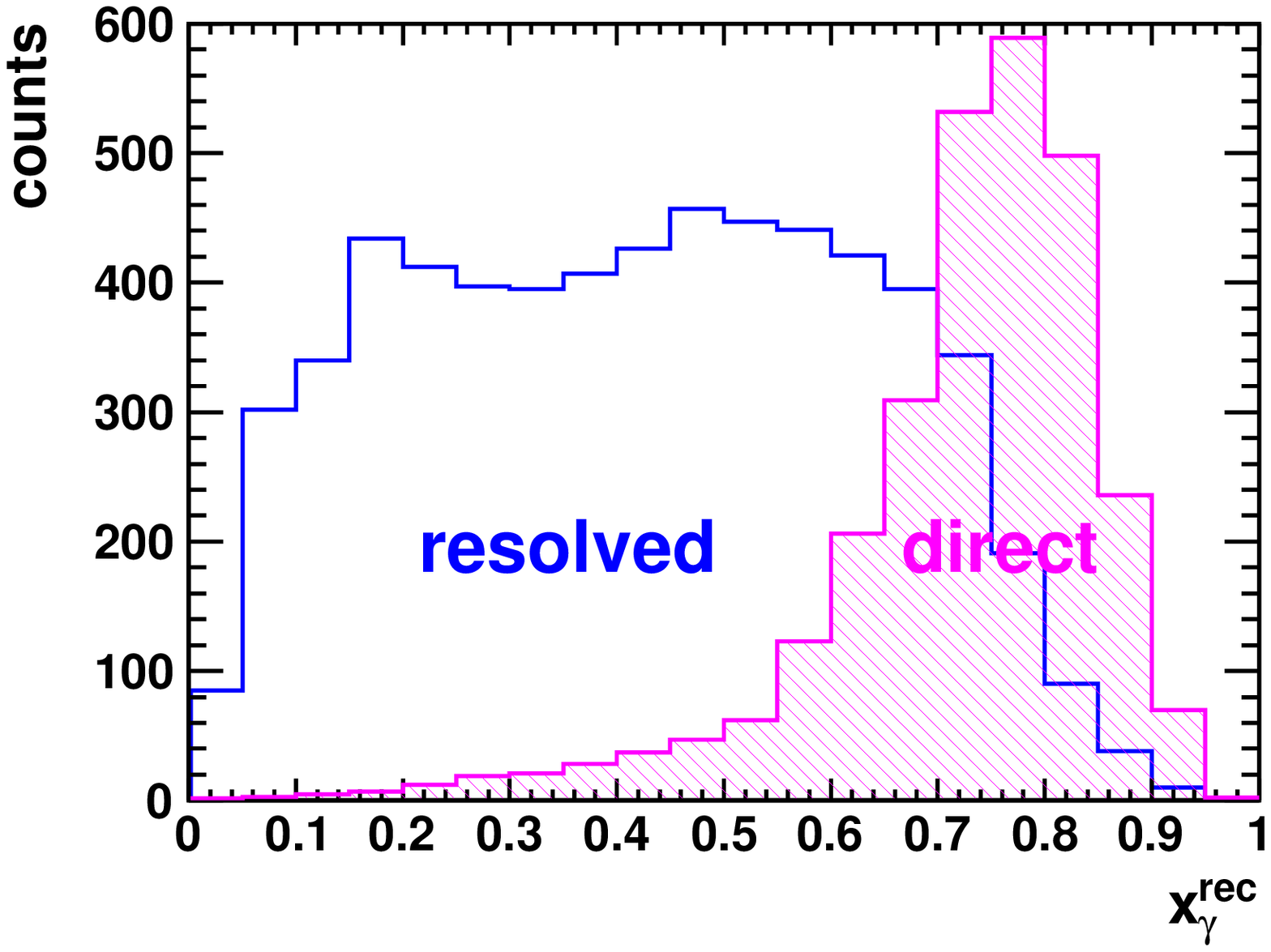}}
\caption{(a) Correlation between $x_{\gamma}^{input}$ and $x_{\gamma}^{rec}$, (b) $x_{\gamma}^{rec}$ distributions in resolved (blank) and direct (shaded) processes.}
\label{}
\end{figure}

The double differential di-jet cross section in photoproduction can be measured with high accuracy in a wide kinematics range of fractional energy $x_{\gamma}^{rec}$ at EIC, see in Fig. 4. In a global fit photon PDFs can be extracted from the cross section.

\begin{figure}[H]
  \centering
  \includegraphics[width=0.4\textwidth]{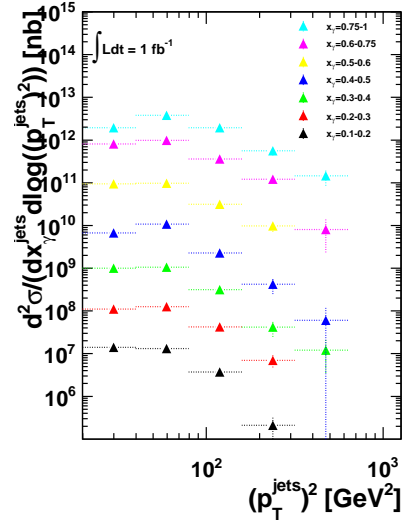}
  \caption{di-jet cross section depends on transverse momentum of the jets $p_{T}^{jets}$ and the reconstructed $x_{\gamma}$ for an integrated luminosity of 1 $fb^{-1}$.}
\end{figure}

\section{Summary and outlook}
In resolved processes, the photon has a hadronic structure. At an EIC, di-jet produced in direct and resolved process can be well separated by reconstructing $x_{\gamma}$. The reconstructed $x_{\gamma}^{rec}$ has a strong correlation with the input $x_{\gamma}$. One can effectively extract the underlying photon PDFs by measuring di-jet cross section in photoproduction events. If we use polarized beams, polarized photon PDFs can be extracted, which is a critical input for ILC $\gamma \gamma$ option.

\section{Acknowledgements}
We are very grateful to the EIC group at BNL whose ongoing efforts made this analysis possible. We acknowledge BNL EIC task force for effective discussions. This work is also supported by the program of Introducing Talents of Discipline to Universities (B08033), the NSFC (11475068) and the National Basic Research Program of China(2013CB837803).

\end{document}